\documentclass{article}
\usepackage{amsfonts}

\usepackage{graphicx}
\usepackage{amsmath}

\input{tcilatex}

\begin{document}

\title{Randomized Hamiltonian Feynman integrals and Schr\"{o}dinger-Ito stochastic
equations}
\author{J. Gough, O. O. Obrezkov, and O. G. Smolyanov}
\date{}
\maketitle

\begin{abstract}
In this paper, we consider stochastic Schr\"{o}dinger equations with
two-dimensional white noise. Such equations are used to describe the
evolution of an open quantum system undergoing a process of continuous
measurement. Representations are obtained for solutions of such equations
using a generalization to the stochastic case of the classical construction
of Feynman path integrals over trajectories in the phase space.
\end{abstract}

AMS 2000 Mathematics Subject Classification. 35A8, 35C15, 35C20, 35R15,
35R60, 46L45, 46N50, 60H15, 60H99, 81Q05, 81Q10, 81Q30, 81S40.

\begin{center}
{\Large Introduction}
\end{center}

In this paper we derive a representation for solutions of Schr\"{o}dinger
equations with white-noise type coefficients (stochastic Schr\"{o}dinger-It%
\={o} equations) using randomized Feynman path integrals over trajectories
in the phase space (Hamiltonian Feynman integrals). These equations describe
the limit behavior of a quantum system observed at discrete instants of time
under the condition that the precision of measurement and the time intervals
between measurements are proportional and tend to zero. Continuous
observation of a quantum system can be defined (informally) as the limit of
these observations. This enables one to assume that Schr\"{o}dinger-It\={o}
equations describe the evolution of a quantum system [20], [22] undergoing a
process of continuous measurement. At the same time, they describe the
so-called Markov approximation for the evolution of an open quantum system%
\footnote{%
By an open quantum system we mean one which is part of a larger quantum
system. The evolution of such a system is described not by the Schr\"{o}%
dinger equation but by a master-equation implied by it. Since
master-equations are integro-differential, it is rather difficult to
investigate them, and therefore use is made of various approximate
equations. The Schr\"{o}dinger-Ito equation can be regarded as one of these.}
as opposed to the approximation given by Hudson Parthasarathy type quantum
stochastic equations (see [12], [4], and [5]).

An equation describing the evolution of a quantum system subject to a
process of continuous measurement for some fixed observable (the operator of
multiplication by a coordinate for an appropriate realization of the Hilbert
state space in the form of $L^{2}(\mathbb{R}^{1}))$ was first postulated in
[25] to describe the spontaneous reduction of the wave function. It was
derived independently by Belavkin [17] in the general situation. (In [25],
use was made of the Hudson Parthasarathy quantum stochastic equations [26];
see also [18].) In the most important special case, this equation was
derived independently by Diosi [21] at the physical level of rigor. For a
derivation on the basis of the standard axioms of quantum mechanics, see [8]
(also [3], [15], and [16]). In the same paper [8] was announced the
stochastic Schr\"{o}dinger equation with two-dimensional white noise. For a
full derivation, see [32]. Apart from the local approach to the description
of the behavior of a continuously observed quantum system resulting in an
evolution equation that generalizes the Schr\"{o}dinger equation and takes
into account the interaction between the quantum system and the measurement
equipment and the effect of this equipment on the state of the system, there
also exists a global approach developed in [30] and [31]. For the global
description of the process of continuous measurement in the latter approach,
a linear propagator for the quantum system is introduced in the form of an
heuristic Feynman path integral over phase-space trajectories.

For the relationship between these two approaches, see [15] and [16].
Representations for solutions of stochastic Schr\"{o}dinger equations via
Feynman path integrals were first obtained in [8], [2] and [16]. In these
papers, the Feynman integral was defined as the analytic continuation of the
integral with respect to the Wiener measure (see [9]), as a result of which
the analytic constraints imposed on the initial condition and on the
potential turned out to be rather restrictive. Furthermore, a representation
for the solution of the stochastic Schr\"{o}dinger equation with
one-dimensional white noise was obtained in [13] under the condition that
the potential and the initial condition in the Cauchy problem are the
Fourier transforms of countably additive measures. In this approach, use is
made of the definition of the Feynman path integral via Parseval's equation
[7], [14], [10], [9]. We use Feynman's original definition [11], [23] of the
functional integral via the limit of finitely multiple integrals and extend
the approach based on Chernoff's theorem [19] to the probabilistic case.
This approach was first used in [35] to obtain a representation for the
solution of the heat equation on a compact Riemannian manifold and then in
[34] for the representation of the solution of the Schr\"{o}dinger equation
using a Feynman path integral over trajectories in the phase space. In the
present paper, we obtain a representation for the solution of the stochastic
Schr\"{o}dinger equation using randomized Feynman path integrals over
trajectories in the phase space (randomized Hamiltonian Feynman integrals).

\section{Pseudo-differential operators and the stochastic Schr\"{o}dinger
equation}

\textbf{Definition 1.} For an arbitrary $\tau \in $[0, 1] we define a map $%
\wedge $: $\mathcal{H}\mapsto \mathcal{\hat{H}}$ from the space of
complex-valued functions on $\mathbb{R}^{1}\times \mathbb{R}^{1}$ to the
space of linear operators on $L^{2}(\mathbb{R})$ as follows: the action of
the operator $\mathcal{\hat{H}}$: $D\left( \mathcal{\hat{H}}\right) \subset
L^{2}\left( R\right) \rightarrow L^{2}\left( R\right) $ on a function $%
\varphi $ is given by the formula (see [34] and [24]) 
\begin{equation}
\left( \mathcal{\hat{H}}\varphi \right) \left( q\right) =\frac{1}{2\pi }%
\lim_{z\rightarrow 0}\int_{-z}^{z}\int_{-z}^{z}\mathcal{H}\left( \left(
1-\tau \right) q+\tau q_{0},p_{0}\right) e^{ip_{0}\left( q-q_{0}\right)
}\varphi \left( q_{0}\right) dq_{0}dp_{0},
\end{equation}
where the limit is taken in $L^{2}\left( \mathbb{R}\right) $. Let $D\left( 
\mathcal{\hat{H}}\right) $ be the set of all $\varphi \in L^{2}\left( 
\mathbb{R}\right) $ such that $\mathcal{\hat{H}}\varphi $ exists.

\bigskip

The function $\mathcal{H}$( $\cdot $ , $\cdot $ ) is called the $\tau $\
-symbol (or the classical Hamiltonian on the phase space $\mathbb{R}%
^{1}\times \mathbb{R}^{1}$) for the pseudo-differential operator $\mathcal{%
\hat{H}}$. The map $\wedge $ determines the $\tau $-quantization.

The operator $\wedge $ for $\tau $= 0 is called the \textit{operator of
qp-quantization} and, for $\tau $= 1, the \textit{operator of pq-quantization%
}. This terminology is related to the fact that the pseudo-differential
operator corresponding to the $\tau $-symbol $H(p,q)=pq$ is equal to $qp$
for $\tau $= 0 and to $pq$ for $\tau $= 1. Here and henceforth, $\hat{q}$
and $\hat{p}$ are the coordinate and momentum operators given by the formulae

\begin{eqnarray*}
\hat{q} &:&f\in Dom\left( \hat{q}\right) \subset L^{2}\left( \mathbb{R}%
\right) \mapsto \left[ q\rightarrow qf\left( q\right) \right] , \\
\hat{p} &:&f\in Dom\left( \hat{p}\right) \subset L^{2}\left( \mathbb{R}%
\right) \mapsto \left[ q\rightarrow -if^{\prime }\left( q\right) \right] ,
\end{eqnarray*}
respectively.

The operation of $\tau $-quantization for $\tau $= 1/2 is called the \textit{%
Weyl quantization}. We also note that if $H(q,p)=f(q)+g(p)$ for all $p,q\in 
\mathbb{R}$ and some functions $f$ and $g$, then the result of quantization
does not depend on the parameter $\tau $.

In what follows, we consider the stochastic Schr\"{o}dinger equation with
two-dimensional white noise and interpret it as the It\={o} stochastic
equation

\begin{eqnarray}
d\varphi \left( t\right)  &=&\left[ \left( -i\mathcal{\hat{H}}-\frac{\mu _{1}%
}{2}k^{2}\left( \hat{q}\right) -\frac{\mu _{2}}{2}h^{2}\left( \hat{p}\right)
\right) \left( \varphi \left( t\right) \right) \right] dt  \notag \\
&&-\sqrt{\mu _{1}}k\left( \hat{q}\right) \left( \varphi \left( t\right)
\right) dW_{1}\left( t\right) -\sqrt{\mu _{2}}h\left( \hat{p}\right) \left(
\varphi \left( t\right) \right) dW_{2}\left( t\right) ,
\end{eqnarray}
where $\mathcal{\hat{H}}$ is the (internal) Hamiltonian obtained for the
observed system by the $\tau $-quantization of the classical Hamiltonian $%
\mathcal{H}$ and where $k(\hat{q})$ and $h(\hat{p})$are the (non-commuting)
differential operators corresponding to the real-valued symbols $%
(q,p)\mapsto 7!k(q)$ and $(q,p)\mapsto h(p)$, respectively. Furthermore, $%
W_{1}$ and $W_{2}$ are independent standard Wiener processes and $\varphi
\left( t\right) \in L^{2}\left( \mathbb{R}\right) $ is a random (wave)
function describing the evolution of mixed states for the observed system.
Equation (2) describes the evolution of an open quantum system undergoing a
continuous measurement of the observables $k(q)$ and $h(p)$. It was
considered in [8] and [32] in the special case when $k(q)=q,h(p)=p$ and $%
H(q,p)=p^{2}/2+V(q)$ for all $p,q\in \mathbb{R}$ and some real-valued
function $V$.

For arbitrary functions $h,k$ and $H$, equation (2) can be obtained by the
method used in [8] and [32]. For this, it suffices to choose an appropriate
realization of the Hilbert state space.

\section{Feynman path integrals over trajectories in the phase space}

Let $E$ be a real vector space, $G\subset E^{\ast }$ the space of linear
functionals on $E$ (it is assumed that $E$ and $G$ satisfy the duality
relation) and $b$ a quadratic functional on $G$. Then, for all $a\in E$ and $%
\alpha \in \mathbb{C}$, the Feynman $\alpha $-pseudo-measure with
correlation functional $b$ and mean value a is a generalized measure $\Phi
_{b,a,\alpha }$\ on $E$ whose Fourier transform is given by the formula

\begin{equation*}
\left( F\Phi _{b,a,\alpha }\right) \left( g\right) =\exp \left\{ \frac{%
\alpha b\left( g\right) }{2}+\alpha g\left( a\right) \right\} ,
\end{equation*}
for all $g\in G$.

Let $Q$ and $P$ be infinite-dimensional locally convex spaces which, as
vector spaces, satisfy the relations $Q=P^{\ast }$ and $P=Q^{\ast }$. If $%
E=Q\times P$ is the phase space and $G=P\times Q$, then the zero-mean
Feynman -pseudo-measure on $E$ with correlation operator given by the
formula $b(p,q)=2p(q)$ for all $(p,q)\in G$ is called the \textit{%
Hamiltonian Feynman integral} (or the Feynman path integral over
trajectories in the phase space).

In this case, the value of the Feynman pseudo-measure on a given function $%
f:Q\times P\mapsto \mathbb{C}$ (the Feynman integral of $f$) can be defined
using the limit of finitely multiple integrals. Let $\left\{ Q_{n}\right\}
_{n}$ and $\left\{ P_{n}\right\} _{n}$ be sequences of finite-dimensional
subspaces (with dim$Q_{n}$ =dim$P_{n}=n$) in $Q$ and $P$, respectively. The
sequential Feynman integral of the function $f$ over trajectories in the
phase space is defined as the limit (if it exists) of the expressions 
\begin{equation*}
\left( \int_{Q_{n}\times P_{n}}e^{ip\left( q\right) }dqdp\right)
^{-1}\int_{Q_{n}\times P_{n}}f\left( q,p\right) e^{ip\left( q\right) }dqdp
\end{equation*}
as $n\rightarrow \infty $, where the integration is with respect to an
arbitrary Lebesgue measure.

For the representation of solutions of Schr\"{o}dinger type equations by
means of sequential Feynman integrals, special importance is attached to the
case of the Feynman path integral over phase-space trajectories for some
spaces $Q$ and $P$ of real-valued functions on the closed interval [0, t].
Let $z\in \mathbb{R}^{2}$, $\tau \in $[0, 1] and $t>0$. Let $Q$ be the
vector space of all real-valued functions on [0, t] whose generalized
derivatives are measures on [0, t] and let $P=\left\{ f\in C\left( \left[ 0,t%
\right] ,\mathbb{R}\right) :f\left( t\right) =0\right\} $. In this case, the
duality relation between $Q$ and $P$ is given by the formula 
\begin{equation*}
\xi _{p}\left( \xi _{q}\right) =\int_{0}^{t}\xi _{p}\left( s\right) \xi
_{q}^{\prime }\left( s\right) ds
\end{equation*}
for all $\xi _{p}\in P$ and $\xi _{q}\in Q$, where $\xi _{q}^{\prime }\left(
s\right) $ denotes the measure equal to the generalized derivative of the
function $\xi _{q}\left( \cdot \right) $.

\bigskip 

\newpage 

\textbf{Definition 2.} The sequential Feynman integral

\begin{equation*}
I(F,z)=\int_{Q\times P^{0}}F\left( \xi _{q},\xi _{p}\right) \Phi ^{\tau
,t,z}\left( d\xi _{q},d\xi _{p}\right)
\end{equation*}
of a function $F:Q\times P^{0}\in \mathbb{C}$ over trajectories in the phase
space $Q\times P^{0}$ is defined as the limit (if it exists) of the finitely
multiple integrals

\begin{eqnarray}
I_{n}(F,z) &=&\frac{1}{\left( 2\pi \right) ^{n}}\int_{\mathbb{R}%
^{2n}}F\left( J_{\tau }\left( q_{0},\cdots ,q_{n}\right) ,J_{1}\left(
p_{0},\cdots ,p_{n}\right) \right)   \notag \\
&&\times \exp \left\{ i\sum_{k=0}^{n-1}p_{k}\left( q_{k+1}-q_{k}\right)
\right\} dq_{0}dp_{0}\cdots dq_{n-1}dp_{n-1}
\end{eqnarray}
as $n\rightarrow \infty $. Here $p_{n}=0$, $q_{n}=z$ and, for each $r\in $%
[0, 1], the expression $J_{r}$ is an (injective) map from $\mathbb{R}^{n+1}$
onto the space consisting of functions that are constant on each of the
intervals $\left( \frac{\left( k-1\right) t}{n},\frac{kt}{n}\right) $, $%
k=1,\cdots ,n$. Furthermore, very $n$-tuple $(q_{0},...,q_{n})\in \mathbb{R}%
^{n+1}$, the function $J_{r}(q_{0},...,q_{n})$ assumes the value $(1-\tau
)q_{k}+\tau q_{k-1}$ on the interval $\left( \frac{\left( k-1\right) t}{n},%
\frac{kt}{n}\right) $, $k=1,\cdots ,n$.

Remark 1 (see [9] and [34]). For $\tau $= 0, the sequential Feynman integral
defined above can be interpreted as an integral over the space of
right-continuous functions. The case $\tau $= 1 corresponds to the
Hamiltonian Feynman integral over the space of left-continuous functions
and, for $\tau \in $(0, 1), the set of functions that forms the convex hull
(with weights (1 - $\tau $) and $\tau $) of the above function spaces should
be regarded as the phase space.

\section{Feynman path integrals and representations of solutions of
evolution equations}

The existence of Feynman path integrals over phase-space trajectories (that
is, the convergence of the corresponding finitely multiple integrals) was
proved in some special cases in [1] and [6] using the finite-difference
method. The definitions of the Feynman integral via the analytic
continuation of the integral with respect to the Wiener measure and via
Parseval's equation were applied in [9] in the representation of solutions
of Schr\"{o}dinger equations.

A new approach based on Chernoff's theorem was first applied in [34]. This
made it possible to extend substantially the area of application of
Feynman's formulae (that is, the representation of solutions of Schr\"{o}%
dinger equations using Feynman integrals). It was noted in [34] that if $%
\varphi :\mathbb{R}_{+}\mapsto L^{2}\left( \mathbb{R}\right) $ is the
solution of the Cauchy problem for the Schr\"{o}dinger equation with initial
datum $\varphi _{0}$ and Hamiltonian equal to the pseudo-differential
operator with $\tau $-symbol $\mathcal{H}$,

\begin{equation}
\frac{d\varphi }{dt}=-i\mathcal{\hat{H}}\varphi ,
\end{equation}
then the relation

\begin{equation}
\varphi \left( t\right) \equiv e^{-it\mathcal{\hat{H}}}\varphi
_{0}=\lim_{n\rightarrow \infty }\left( \widehat{e^{-i\frac{t}{n}\mathcal{H}}}%
\right) ^{n}\varphi _{0}
\end{equation}
is a representation for the solution of the Schr\"{o}dinger equation via the
Feynman path integral over trajectories in the phase space. Indeed, it can
be verified that the right-hand side of (5) is a function whose value at a
point $z$ coincides with the limit of finitely multiple approximations of
the Feynman integral

\begin{equation*}
\int \exp \left\{ -i\int_{0}^{t}\mathcal{H}\left( \xi _{q}\left( s\right)
,\xi _{p}\left( s\right) \right) ds\right\} \varphi _{0}\left( \xi
_{q}\left( 0\right) \right) \Phi ^{\tau ,t,z}\left( d\xi _{q},d\xi
_{p}\right) .
\end{equation*}
The formula (5) was proved in [34] for a rather wide class of Hamiltonians
using Chernoff's theorem.

In the present paper, we extend this approach to stochastic differential
equations of type (2). In this case, the corresponding Feynman formula
changes. We shall show that if a random function $\varphi :\mathbb{R}%
_{+}\mapsto $ $L^{2}\left( \mathbb{R}\right) $ is the solution of the Cauchy
problem for the stochastic equation

\begin{equation}
d\varphi =\hat{A}\varphi dt+\hat{B}\varphi dW\left( t\right)
\end{equation}
with initial datum $\varphi $, where $\hat{A}$ and $\hat{B}$ are
pseudo-differential operators on $L^{2}\left( \mathbb{R}\right) $ and $W$ is
the standard Wiener process, then, under certain conditions on $A$ and $B$,
the relation

\begin{equation}
\varphi (t)=\lim_{n\rightarrow \infty }\text{hat}\left( \exp \left\{ -\frac{%
tB^{2}}{2n}+B\Delta W_{k,n}+\frac{t}{n}A\right\} \right) \varphi _{0}
\end{equation}
holds, where $\Delta W_{k,n}=W(tk/n)-W(t(k-1)/n)$ for all $k$, $k=1,...,n$,
and hat$(M)=\hat{M}$. The additional factors $\exp \left\{ -\frac{tB^{2}}{2n}%
\right\} $ under the product sign correspond to It\={o}'s formula. The
right-hand side of (7) can be interpreted as a randomized Hamiltonian
Feynman integral.

\section{A stochastic analogue of Feynman's formula}

The approach used in [34] to find solutions of non-stochastic Schr\"{o}%
dinger equations is based on the construction of a family of operators
approximating in the sense of Chernoff (see [36]) the resolvent operator
semigroup for the Schr\"{o}dinger equation. Let $D_{1}$ be an essential
domain for the operator $\mathcal{\hat{H}}$, which means that the operator $%
\left( \mathcal{\hat{H}},D\left( \mathcal{\hat{H}}\right) \right) $ is the
closure of $\left( \mathcal{\hat{H}},D_{1}\right) $. The one-parameter
operator family $\{S(t)\}_{t>0}$ approximates the semigroup with generator $%
-i\mathcal{\hat{H}}$ (by definition, it is precisely the resolvent semigroup
of equation (2) for $%
{\mu}%
_{1}=%
{\mu}%
_{2}=0$) in the sense of Chernoff if the relation $S(t)f=f-it\mathcal{\hat{H}%
}f+o(t),$ $t\rightarrow 0,$holds for all $f\in D_{1}$. Then Chernoff's
theorem implies that the relation

\begin{equation*}
e^{-it\mathcal{\hat{H}}}\varphi _{0}=\lim_{n\rightarrow \infty }\left(
S\left( \frac{t}{n}\right) \right) ^{n}\varphi _{0}
\end{equation*}
holds for all $t>0$ and $\varphi _{0}\in L^{2}\left( \mathbb{R}\right) $. If 
$\widehat{e^{-it\mathcal{H}}}$ is taken as $S(t)$, then we obtain the
representation (5).

This method cannot be used for the representation of solutions in the case
of stochastic Schr\"{o}dinger type equations (with $%
{\mu}%
_{1},%
{\mu}%
_{2}>0$ in (2)) since the solution is a random function in $L^{2}\left( 
\mathbb{R}\right) $ and the family of operators approximating the resolvent
family for equation (2) is non-deterministic. We also note that, since the
right-hand side involves a random process, the semigroup property does not
hold for the resolvent family corresponding to the stochastic equation.
Nevertheless, if $\left\{ T_{r}^{s}\right\} _{s\geq r\geq 0}$ is the
resolvent family of (random) operators on $L^{2}\left( \mathbb{R}\right) $
that corresponds to the Cauchy problem for the stochastic equation (2) (that
is, $T_{r}^{s}$ is defined by the formula $T_{r}^{s}\varphi _{0}=\varphi
\left( s\right) $ for arbitrary $s$ and $r,s\geq r\geq 0$, where $\varphi $
is the solution of (2) with $\varphi \left( r\right) =\varphi _{0}$, under
the assumption that such a solution exists and is unique), then, since $%
W_{1} $ and $W_{2}$ are processes with independent increments, we have the
following stochastic semigroup property: the distribution of $T_{r}^{s}$
depends only on the difference $s-r$. This enables us to generalize the
approach based on the notion of equivalence in the sense of Chernoff to the
stochastic case.

In what follows, we consider equation (2) for the case $\tau $= 0, that is,
the operator $\mathcal{\hat{H}}$ is obtained from $\mathcal{H}$ by means of
qp-quantization. It is assumed that the formula (5) holds for the
Hamiltonian $\mathcal{\hat{H}}$. A sufficient condition [34] for the
fulfillment of (5) is that the relation 
\begin{equation}
H(q,p)=k_{0}(q)+h_{0}(p)+l(q,p)
\end{equation}
hold for all $q,p\in \mathbb{R}$ and some real-valued functions $%
k_{0},h_{0},l\in L^{2}\left( \mathbb{R}\right) $. In addition, we assume
that $-i\mathcal{\hat{H}}$ is the generator of a strongly continuous
operator semigroup.

We consider the family $\left\{ Q_{r}^{s}\right\} _{0\leq r\leq s}$ of
random operators on $L^{2}\left( \mathbb{R}\right) $ defined by the formula

\begin{equation}
(Q_{r}^{s}f)\left( q\right) =\exp \left\{ -\sqrt{\mu _{1}}k\left( q\right)
\left( W_{1}\left( s\right) -W_{1}\left( r\right) \right) -\mu _{1}\left(
s-r\right) k^{2}\left( q\right) \right\} f\left( q\right) ,\quad q\in 
\mathbb{R}.
\end{equation}
We claim that the function $s\mapsto Q_{r}^{s}f$ is the solution of the
Cauchy problem for the equation

\begin{equation}
d\varphi \left( t\right) =-\frac{\mu _{1}}{2}k^{2}\left( \hat{q}\right)
\left( \varphi \left( t\right) \right) dt-\sqrt{\mu _{1}}k\left( \hat{q}%
\right) \left( \varphi \left( t\right) \right) dW_{1}\left( t\right)
\end{equation}
with initial condition $\varphi (r)=f$.

We take the total derivative of $Q_{r}^{s}f$ with respect to $s$,

\begin{eqnarray*}
d_{s}(Q_{r}^{s}f)(q) &=&\{-\sqrt{\mu _{1}k\left( q\right) }dW_{1}\left(
s\right) +\frac{1}{2}\mu _{1}k^{2}\left( q\right) \left( dW_{1}\left(
s\right) \right) ^{2} \\
&&-\mu _{1}\left( s-r\right) k^{2}\left( q\right) ds\}\left(
Q_{r}^{s}f\right) \left( q\right) .
\end{eqnarray*}
According to It\={o}'s formula $\left( dW_1(s)\right) ^{2}=ds$, it follows
that $s\mapsto Q_{r}^{s}f$ is the solution of equation (10). The fulfillment
of the initial condition is obvious.

It can be similarly shown that if $\left\{ P_{r}^{s}\right\} _{0\leq r\leq
s} $ is the operator family given by the formula 
\begin{multline}
\left( P_{r}^{s}f\right) \left( q\right) =\frac{1}{2\pi }\int_{\mathbb{R}%
}\int_{\mathbb{R}}\exp \left\{ -\sqrt{\mu _{2}}h\left( p_{0}\right) \left(
W_{2}\left( s\right) -W_{2}\left( r\right) \right) -\mu _{2}\left(
s-r\right) h^{2}\left( p_{0}\right) \right\}  \notag \\
\times e^{ip_{0}\left( q-q_{0}\right) }dq_{0}dp_{0},\quad q\in \mathbb{R},
\end{multline}
then $s\mapsto P_{r}^{s}f$ is the solution of the Cauchy problem for the
equation

\begin{equation}
d\varphi \left( t\right) =-\frac{\mu _{2}}{2}h^{2}\left( \hat{p}\right)
\left( \varphi \left( t\right) \right) dt-\sqrt{\mu _{2}}h^{2}\left( \hat{p}%
\right) \left( \varphi \left( t\right) \right) dW_{2}\left( t\right)
\end{equation}
with initial condition $\varphi (r)=f$.

As shown in [34], if $Y_{t}=\widehat{e^{-it\mathcal{H}}}$ for all $t$, then
the function $t\mapsto Y_{t}\varphi _{0}$ approximates the solution of the
equation

\begin{equation}
d\varphi \left( t\right) =-i\mathcal{\hat{H}}\varphi \left( t\right) dt
\end{equation}
in the sense of Chernoff.

We note that the sum of the right-hand sides of the (linear) equations (10),
(12) and (13) coincides with the right-hand side of (2), and it is therefore
to be expected that the operator family $\left\{ U_{r}^{s}\right\} _{0\leq
r\leq s}$ determined by the formula $U_{r}^{s}=Q_{r}^{s}Y_{s-r}P_{r}^{s}$
approximates, in a sense, the resolvent family for (2).

The following relation will be called the \textit{Feynman stochastic formula}%
: 
\begin{equation}
T_{0}^{t}\varphi _{0}=w-\lim_{n\rightarrow \infty }U_{\left( n-1\right)
t/n}^{t}\cdots U_{0}^{t/n}\varphi _{0},
\end{equation}
where $w-\lim $ denotes a kind of convergence (defined below) for $%
L^{2}\left( \mathbb{R}\right) $-valued random variables. To see an analogy
with the usual Feynman formula, it suffices to note that $t\mapsto
T_{0}^{t}\varphi _{0}$ is the solution of equation (2) with initial datum $%
\varphi _{0}$ and that the right-hand side of (14) is a finite-dimensional
approximation to the Feynman integral of a random function (see (7)).
Finally, for $\mu _{1}=\mu _{2}=0$, we obtain the standard Feynman formula
in [34].

\bigskip

\textbf{Definition 3.} Let $\left\{ \xi _{n}\right\} $ be random variables
defined on a probability space $(\Omega ,\mathcal{G},\mathbb{P})$ and having
values in $L^{2}\left( \mathbb{R}\right) $. A random variable $\xi $\ on the
same space is the $w$-limit of $\xi _{n}$ as $n\rightarrow \infty $ if and
only if $\mathbb{E}\left\| \xi -\xi _{n}\right\| _{L^{2}\left( \mathbb{R}%
\right) }^{2}$ tends to zero as $n\rightarrow \infty $. Here $\mathbb{E}$
denotes the operator of mathematical expectation in the probability space $%
(\Omega ,\mathcal{G},\mathbb{P})$.

\bigskip

To prove formula (14), it is necessary to consider asymptotic properties of
the operator $U_{r}^{s}$ as $s-r\rightarrow 0$. In what follows, we use a
well-known result of L\'{e}vy [28] (see also [29]) on the local smoothness
of a Wiener process.

\bigskip

\textbf{Lemma 1}. Let $\{W(r)\}_{0\leq r\leq 1}$ be the standard Wiener
process. Then the relation 
\begin{equation*}
\overline{\lim_{v\rightarrow 0}}\sup_{r\in \left[ 0,1\right] }\frac{|W\left(
r+v\right) -W\left( r\right) |}{\sqrt{-2v\ln v}}=1
\end{equation*}
holds with probability 1.

\bigskip

It follows from Lemma 1 that the relation 
\begin{equation*}
\overline{\lim_{v\rightarrow 0}}\sup_{r\in \left[ 0,1\right] }\frac{|W\left(
r+v\right) -W\left( r\right) |}{v^{\frac{1}{2}-\varepsilon }}=0
\end{equation*}
holds for each $\varepsilon $ \TEXTsymbol{>} 0 with probability 1.

For $q\in \mathbb{R}$ we write $\varepsilon _{2}\left( q\right) =e^{q}-1-q-%
\frac{q^{2}}{2}$. It is clear that there is a constant $A_{1}$\TEXTsymbol{>}
0 such that the inequality \TEXTsymbol{\vert}$\varepsilon _{q}\left(
q\right) $\TEXTsymbol{\vert}$\leq A_{1}|q|^{3}$ holds as $q\rightarrow 0$.
For brevity, in what follows we shall use the notation $\Delta
W_{j}^{r,v}=W_{j}\left( r+v\right) -W_{j}\left( r\right) $ for $j=1,2$, and
therefore the relation

\begin{multline*}
\exp \left\{ -\sqrt{\mu _{1}}k\left( q\right) \left( W_{1}\left( r+v\right)
-W_{1}\left( r\right) \right) -\mu _{1}vk^{2}\left( q\right) \right\}  \\
=1-\sqrt{\mu _{1}}k\left( q\right) \Delta W_{1}^{r,v}-\mu _{1}vk^{2}\left(
q\right)  \\
+\frac{1}{2}\left[ \sqrt{\mu _{1}}\Delta W_{1}^{r,v}k\left( q\right) +\mu
_{1}vk^{2}\left( q\right) \right] ^{2}+\varepsilon _{2}\left( \sqrt{\mu _{1}}%
\Delta W_{1}^{r,v}k\left( q\right) +\mu _{1}vk^{2}\left( q\right) \right)  \\
\equiv 1-\sqrt{\mu _{1}}k\left( q\right) \Delta W_{1}^{r,v}-\mu
_{1}vk^{2}\left( q\right) +\frac{1}{2}\left[ \sqrt{\mu _{1}}\Delta
W_{1}^{r,v}k\left( q\right) +\mu _{1}vk^{2}\left( q\right) \right]
^{2}+c\left( v,r,q\right) 
\end{multline*}
will hold for the function defining the action of the operator $Q_{r}^{r+v}$
(see (9)).

It is assumed below that the function $k$ is bounded ($\sup_{q\in \mathbb{R}%
}|k(q)|=K1<\infty $) and belongs to $L^{2}\left( \mathbb{R}\right) $. We
claim that the relation $\mathbb{E}\left\| c(v,r,\cdot )\right\|
_{L^{2}\left( \mathbb{R}\right) }=O\left( v^{3/2}\right) $ then holds
uniformly with respect to $r$ as $v\rightarrow 0$. By definition,

\begin{eqnarray*}
c(v,r,q) &=&\frac{1}{2}\mu _{1}^{2}v^{2}k^{4}\left( q\right) +\left( \mu
_{1}\right) ^{3/2}v\Delta W_{1}^{r,v}k^{3}\left( q\right)  \\
&&+\varepsilon _{2}\left( \sqrt{\mu _{1}}\Delta W_{1}^{r,v}k\left( q\right)
+\mu _{1}vk^{2}\left( q\right) \right) .
\end{eqnarray*}
Therefore we have the inequalities

\begin{gather*}
\left\| v^{2}k^{4}\right\| _{L^{2}}\leq \left\| k\right\|
_{L^{2}}K_{1}^{3}v^{2}=O\left( v^{3/2}\right) , \\
\mathbb{E}\left\| v\Delta W_{1}^{r,v}k^{3}\right\| _{L^{2}}\leq \left\|
k\right\| _{L^{2}}K_{1}^{2}v\mathbb{E}|\Delta W_{1}^{r,v}| \\
=\left\| k\right\| _{L^{2}}K_{1}^{2}v\frac{1}{\sqrt{2\pi v}}\int_{\mathbb{R}%
}|z|e^{-z^{2}/(2v)}dz \\
=\left\| k\right\| _{L^{2}}K_{1}^{2}v\frac{2}{\sqrt{2\pi }}\sqrt{v}=O\left(
v^{3/2}\right) , \\
\mathbb{E}\left\| \varepsilon _{2}\left( \sqrt{\mu _{1}}\Delta
W_{1}^{r,v}k\left( q\right) +\mu _{1}vk^{2}\left( q\right) \right) \right\|
_{L^{2}}\leq \\
A_{1}\mathbb{E}\left\| \varepsilon _{2}\left( \sqrt{\mu _{1}}\Delta
W_{1}^{r,v}k\left( q\right) +\mu _{1}vk^{2}\left( q\right) \right)
^{3}\right\| _{L^{2}}
\end{gather*}

The expression on the right-hand side of the third of these inequalities has
an upper bound equal to the sum $\sum_{j=0}^{3}b_{j}\mathbb{E}|\Delta
W_{1}^{r,v}|^{j}\left\| k^{j}\right\| _{L^{2}}v^{3-j}\left\| k^{2\left(
3-j\right) }\right\| _{L^{2}}$ with some constants $b_{j}$. Since $\mathbb{E}%
|\Delta W_{1}^{r,v}|^{j}$ is proportional to $v^{j/2}$ and we have $\left\|
k\alpha \right\| _{L^{2}}\leq \left\| k\right\| _{L^{2}}K_{1}^{\alpha -1}$
for all $\alpha $\ \TEXTsymbol{>} 1, each of the terms in this sum is of
order $O(v^{3/2})$. Consequently, $\mathbb{E}\left\| c\left( v,r,\cdot
\right) \right\| _{L^{2}}=O(v^{3/2})$, that is, $\mathbb{E}\left\| c\left(
v,r,\cdot \right) \right\| _{\mathcal{L}\left( L^{2}\right) }=O(v^{3/2})$,
where $\mathcal{L}\left( L^{2}\right) $ is the space of continuous linear
operators in $L^{2}\left( \mathbb{R}\right) $. The uniformity of the
estimates in $r$ follows from the fact that the distribution of $\Delta
W_{1}^{r,v}$ does not depend on $r$.

Everywhere below, $\mathcal{Q}_{r}^{s}$ with arbitrary $s$ and $r,s>r>0$,
denotes the minimal $\sigma $-algebra relative to which the random variables 
$\{W_{1}(t)-W_{1}(r)\}$ are measurable. We note that, since $W_{1}$is a
process with independent increments, the $\sigma $-algebras $\mathcal{Q}%
_{r_{1}}^{s_{1}}$ and $\mathcal{Q}_{r_{2}}^{s_{2}}$ are independent if the
intervals $(r_{1},s_{1})$ and $\left( r_{2},s_{2}\right) $ are disjoint.

\bigskip

\textbf{Lemma 2}. Let $k$ be a bounded function belonging to $L^{2}\left( 
\mathbb{R}\right) $. Then for arbitrary $r,v>0$, we have the following
asymptotic expansion of the random operator $Q_{r}^{r+v}$ with respect to
the parameter $v$:

\begin{equation}
Q_{r}^{r+v}=1-\sqrt{\mu _{1}}k\left( \hat{q}\right) \Delta W_{1}^{r,v}-\frac{%
\mu _{1}}{2}k^{2}\left( \hat{q}\right) +\frac{\mu _{1}}{2}vk^{2}\left( \hat{q%
}\right) \zeta ^{r,v}+O\left( v^{3/2}\right) ,\quad v\rightarrow 0,
\end{equation}
where $O(v^{3/2})$ denotes a random operator in $L^{2}\left( \mathbb{R}%
\right) $ measurable relative to $\mathcal{Q}_{r}^{r+v}$ such that the
mathematical expectation of its norm is of order $O(v^{3/2})$ uniformly with
respect to $r$. Furthermore, $\zeta ^{r,v}$ is a zero-mean random variable
measurable relative to $\mathcal{Q}_{r}^{r+v}$ whose distribution does not
depend on $r$ or $v$, and 1 denotes the identity operator in $L^{2}\left( 
\mathbb{R}\right) $.

\textit{Proof.} It follows from the previous argument that

\begin{eqnarray*}
Q_{r}^{r+v} &=&1-\sqrt{\mu _{1}}k\left( \hat{q}\right) \Delta
W_{1}^{r,v}-\mu _{1}vk^{2}+\frac{\mu _{1}}{2}k^{2}\left( \Delta
W_{1}^{r,v}\right) ^{2}+O\left( v^{3/2}\right) \\
&=&1-\sqrt{\mu _{1}}k\left( \hat{q}\right) \Delta W_{1}^{r,v}-\mu _{1}vk^{2}+%
\frac{\mu _{1}}{2}vk^{2}\frac{\left( \Delta W_{1}^{r,v}\right) ^{2}-v}{v}%
+O\left( v^{3/2}\right) .
\end{eqnarray*}
It remains to note that $\zeta ^{r,v}=\frac{\left( \Delta W_{1}^{r,v}\right)
^{2}-v}{v}$ is a zero-mean random variable distributed according to the law $%
\chi ^{2}\left( 1\right) -1$.

An asymptotic expansion for $P_{r}^{r+v}$ can be obtained in a similar way
(everywhere below, $\mathcal{P}_{r}^{s}$ denotes the minimal $\sigma $%
-algebra relative to which the random variables \{$W_{2}(t)-W_{2}(r)\}_{r%
\leq t\leq s}$ are measurable). If $h\in L^{2}\left( \mathbb{R}\right) $ and 
$\sup_{q\in \mathbb{R}}|h(q)|<1$, then the relation 
\begin{equation}
P_{r}^{r+v}=1-\sqrt{\mu _{2}}h\left( \hat{p}\right) \Delta W_{2}^{r,v}-\frac{%
\mu _{2}}{2}vh^{2}\left( \hat{p}\right) +\frac{\mu _{2}}{2}vh^{2}\left( \hat{%
p}\right) \eta ^{r,v}+O\left( v^{3/2}\right) ,\quad v\rightarrow 0,
\end{equation}
holds, where $O(v^{3/2})$ denotes a random operator in $L^{2}\left( \mathbb{R%
}\right) $ measurable relative to $\mathcal{P}_{r}^{r+v}$ such that the
mathematical expectation of its norm is of order $O(v^{3/2})$ uniformly with
respect to $r$. Furthermore, $\eta ^{r,v}$ is a zero-mean random variable
measurable relative to $\mathcal{P}_{r}^{r+v}$ whose distribution does not
depend on $r$ or $v$.

To prove (16), it suffices to note that the operator $P_{r}^{r+v}$ \ can be
written in the using the Fourier transform $F$ in $L^{2}$: 
\begin{equation*}
P_{r}^{r+v}=F\exp \left\{ -\sqrt{\mu _{2}}h\Delta W_{2}^{r,v}-\mu
_{2}vh^{2}\right\} F^{-1}.
\end{equation*}
An expansion similar to (15) can be obtained for the exponential in this
formula, and the application of the Fourier transform operator on the right
and left of the exponential leads to the asymptotic expansion (16) since
this operator in unitary and non-random.

The asymptotic expansions (15) and (16), the definition of the Hamiltonian
symbol (8) and Lemma 1 imply an asymptotic expansion for the operator $%
U_{r}^{r+v}=Q_{r}^{r+v}\widehat{e^{-iv\mathcal{H}}}P_{r}^{r+v}$: 
\begin{eqnarray*}
U_{r}^{r+v} &=&1-\sqrt{\mu _{1}}k\left( \hat{q}\right) \Delta W_{1}^{r,v}-%
\frac{\mu _{1}}{2}vk^{2}\left( \hat{q}\right) +\frac{\mu _{1}}{2}%
vk^{2}\left( \hat{q}\right) \zeta ^{r,v}-\sqrt{\mu _{2}}h\left( \hat{p}%
\right) \Delta W_{2}^{r,v} \\
&&-\frac{\mu _{2}}{2}vh^{2}\left( \hat{p}\right) +\frac{\mu _{2}}{2}%
vh^{2}\left( \hat{p}\right) \eta ^{r,v}-iv\left( k_{0}\left( \hat{q}\right)
+h_{0}\left( \hat{p}\right) +\hat{l}\right) \\
&&+\sqrt{\mu _{1}\mu _{2}}k\left( \hat{q}\right) h\left( \hat{p}\right)
\Delta W_{1}^{r,v}\Delta W_{2}^{r,v}+o\left( v^{3/2-\varepsilon }\right) ,
\end{eqnarray*}
where $\varepsilon $\TEXTsymbol{>} 0 is an arbitrary number.

To derive a similar asymptotic expansion for the operator $T_{r}^{r+v}$, we
need the auxiliary assertion below.

\bigskip

\textbf{Lemma 3}. Let $A$ and $B$ be operators in a Banach space $X$, let $W$
denote the standard Wiener process and let $\varphi :\mathbb{R}_{+}\mapsto X$
be a random function satisfying the It\={o} stochastic equation $d\varphi
=A\varphi dt+B\varphi dW\left( t\right) $. Then for any $\varepsilon $ 
\TEXTsymbol{>} 0, the asymptotic expansion

\begin{equation*}
\varphi (t)=\varphi (0)+B\varphi (0)W(t)+A\varphi (0)t+\frac{1}{2}%
B^{2}\varphi \left( 0\right) \left( W\left( t\right) ^{2}-t\right) +o\left(
t^{\frac{3}{2}-\varepsilon }\right) ,\quad t\rightarrow 0,
\end{equation*}
holds. Here $o\left( t^{\frac{3}{2}-\varepsilon }\right) $ denotes a random
variable $\xi _{t}$ such that the relation

\begin{equation*}
\lim_{t\rightarrow 0}\frac{|\xi _{t}|}{t^{\frac{3}{2}-\varepsilon }}=0
\end{equation*}
holds with probability 1.

\textit{Proof.} By the definition of a solution of the It\={o} stochastic
differential equation, the function $\varphi $ satisfies the integral
relation

\begin{equation}
\varphi \left( t\right) -\varphi \left( 0\right) =\int_{0}^{t}A\varphi
\left( r\right) dr+\int_{0}^{t}B\varphi \left( r\right) dW(r),
\end{equation}
where $\int_{0}^{t}B\varphi \left( r\right) dW(r)$ is the It\={o} stochastic
integral. We shall find an asymptotic expansion for $\varphi \left( t\right) 
$ as $t\rightarrow 0$ using the method of indeterminate coefficients. Let 
\begin{equation*}
\varphi (t)=\varphi (0)+\beta W(t)+\alpha t+\frac{1}{2}\gamma \left( W\left(
t\right) ^{2}-t\right) +o\left( t^{\frac{3}{2}-\varepsilon }\right) .
\end{equation*}
We substitute the right-hand side of this equation into the right- and
left-hand sides of equation (18). Since we are interested in an asymptotic
expansion to within $o\left( t^{\frac{3}{2}-\varepsilon }\right) $, we can
take into account only the constant in the asymptotic expansion obtained
when $\varphi \left( r\right) $ is substituted in $\int_{0}^{t}A\varphi
\left( r\right) dr$ (by Lemma 1, the other terms are of order $o\left( r^{%
\frac{1}{2}-\varepsilon }\right) $). Consequently, 
\begin{equation*}
\int_{0}^{t}A\varphi \left( r\right) dr=A\varphi \left( 0\right) t+o\left(
t^{\frac{3}{2}-\varepsilon }\right) .
\end{equation*}
We similarly conclude that 
\begin{eqnarray*}
\int_{0}^{t}B\varphi \left( r\right) dW(r) &=&B\varphi \left( 0\right)
W\left( t\right) +\int_{0}^{t}B\beta W\left( r\right) dW(r)+o\left( t^{\frac{%
3}{2}-\varepsilon }\right)  \\
&=&B\varphi \left( 0\right) W\left( t\right) +\frac{1}{2}B\beta \left[
W\left( t\right) ^{2}-1\right] +o\left( t^{\frac{3}{2}-\varepsilon }\right) .
\end{eqnarray*}
Equating the left- and right-hand sides of (18), we arrive at the formula

\begin{eqnarray*}
&&\beta W\left( t\right) +\alpha t+\frac{\gamma }{2}\left[ W\left( t\right)
^{2}-1\right] +o\left( t^{\frac{3}{2}-\varepsilon }\right) \\
&=&A\varphi \left( 0\right) t+B\varphi \left( 0\right) W\left( t\right) +%
\frac{1}{2}B\beta \left[ W\left( t\right) ^{2}-1\right] +o\left( t^{\frac{3}{%
2}-\varepsilon }\right) ,
\end{eqnarray*}
whence we obtain the relations $\alpha =A\varphi \left( 0\right) $, $\beta
=B\varphi \left( 0\right) $ and $\gamma =B\beta $, and the assertion of the
lemma follows.

Using Lemma 3 in the case of equation (2), we derive an asymptotic expansion
for the operator $T_{r}^{r+v}$ :

\begin{eqnarray*}
T_{r}^{r+v} &=&1-\sqrt{\mu _{1}}k\left( \hat{q}\right) \Delta W_{1}^{r,v}-%
\frac{\mu _{1}}{2}vk^{2}\left( \hat{q}\right)  \\
&&+\frac{\mu _{1}}{2}vk^{2}\left( \hat{q}\right) \zeta ^{r,v}-\sqrt{\mu _{2}}%
h\left( \hat{p}\right) \Delta W_{2}^{r,v}-\frac{\mu _{2}}{2}vh^{2}\left( 
\hat{p}\right)  \\
&&+\frac{\mu _{2}}{2}vh^{2}\left( \hat{p}\right) \eta ^{r,v}-iv\left(
k_{0}\left( \hat{q}\right) +h_{0}\left( \hat{p}\right) +\hat{l}\right)
+o\left( v^{3/2-\varepsilon }\right) ,
\end{eqnarray*}
where, as usual, $\xi ^{r,v}=\frac{\left( \Delta W_{1}^{r,v}\right) ^{2}-v}{v%
}$ and $\eta ^{r,v}=\frac{\left( \Delta W_{2}^{r,v}\right) ^{2}-v}{v}$. We
note that the direct application of Lemma 3 proves the expansion (19) only
for $r=0$. However, since $W_{1}$ and $W_{2}$ are Wiener processes, the
distribution of the operator-valued random variable $T_{r}^{r+v}$ does not
depend on $r$, and the resulting expansion holds for an any $r$. Moreover,
the expression $o\left( v^{3/2-\varepsilon }\right) $ on the right-hand side
of (19) is an operator in $L^{2}\left( \mathbb{R}\right) $ such that the
mathematical expectation of its norm is of order $o\left( v^{3/2-\varepsilon
}\right) $ uniformly with respect to $r$.

As can be seen from the asymptotic expansions (19) and (17), the difference
between $T_{r}^{r+v}$ and $U_{r}^{r+v}$ is equal to the term $\sqrt{\mu
_{1}\mu _{2}}k\left( \hat{q}\right) h\left( \hat{p}\right) \Delta
W_{1}^{r,v}\Delta W_{2}^{r,v}+o\left( v^{3/2-\varepsilon }\right) $ which
plays a key role in the proof of the randomized Feynman formula (14). We
also note that, in its standard form, the Chernoff theorem used in the
derivation of Feynman type formulae involves, apart from the approximation
requirement for the operator family, the condition that the norm $\left\|
S(t)\right\| $ of the operator in the approximating family should not exceed 
$e^{Ct}$, where $C>0$ is some constant (common to all values of $t>0$).

It turns out that a similar technical constraint is also necessary for
proving the randomized Feynman formula.

\bigskip

\textbf{Lemma 4}. Let $h$ and $k$ be bounded functions belonging to $%
L^{2}\left( \mathbb{R}\right) $. Let the map $t\mapsto \mathbb{E}\left\|
T_{0}^{t}\right\| $ be differentiable at zero. Then there is a constant C 
\TEXTsymbol{>} 0 such that, for all $r,v>0$, the estimates 
\begin{equation*}
\mathbb{E}\left\| U_{r}^{r+v}\right\| \leq e^{Cv},\mathbb{E}\left\|
T_{r}^{r+v}\right\| \leq e^{Cv}
\end{equation*}
hold, where$\left\| \cdot \right\| $ denotes the norm on the space of linear
operators in $L^{2}\left( \mathbb{R}\right) $.

\textit{Proof.} It suffices to consider the case $r=0$ (the estimates for
arbitrary values of $r>0$ will be the same as those for$r=0$ since the
distributions of the operator-valued random variables under consideration
depend only on the difference between the superscript and the subscript). We
first derive an estimate for $U_{0}^{v}$. If $K_{1}=\sup_{q\in \mathbb{R}%
}|k\left( q\right) |<\infty $, then $\left\| Q_{0}^{v}\right\| \leq e^{K_{1}%
\sqrt{\mu _{1}}\Delta W_{1}\left( v\right) }$. Therefore $\mathbb{E}\left\|
Q_{0}^{v}\right\| \leq e^{\frac{1}{2}K_{1}^{2}\mu _{1}v}$.

Similarly, if $K_{2}=\sup_{q\in \mathbb{R}}|k\left( q\right) |<\infty $,
then $\mathbb{E}\left\| P_{0}^{v}\right\| \leq e^{\frac{1}{2}K_{2}^{2}\mu
_{2}v}$. To see this, it suffices to note that, by the definition of the
quantization operator, we have 
\begin{equation*}
P_{0}^{v}=F^{-1}\exp \left\{ \sqrt{\mu _{2}}h\left( \cdot \right)
W_{2}\left( v\right) -\mu _{2}vh^{2}\left( \cdot \right) \right\} F,
\end{equation*}
where $\exp \left\{ \sqrt{\mu _{2}}h\left( \cdot \right) W_{2}\left(
v\right) -\mu _{2}vh^{2}\left( \cdot \right) \right\} $ is the operator of
multiplication by the corresponding function in $L^{2}\left( \mathbb{R}%
\right) $ and $F$ is the Fourier transform operator in $L^{2}\left( \mathbb{R%
}\right) $. Then the inequality $\mathbb{E}\left\| P_{0}^{v}\right\| \leq e^{%
\frac{1}{2}K_{2}^{2}\mu _{2}v}$ follows from the fact that the Fourier
transform operator is unitary along with the estimate for the mathematical
expectation of the norm of the operator $\exp \left\{ \sqrt{\mu _{2}}h\left(
\cdot \right) W_{2}\left( v\right) -\mu _{2}vh^{2}\left( \cdot \right)
\right\} $.

We note that, since the operator-valued random variable $Q_{0}^{v}\;\left(
P_{0}^{v}\right) $ depends only on the realization of the process $W_{1}$ ($%
W_{2}$) respectively and the processes $W_{1}$ and $W_{2}$ are independent,
we have the relation 
\begin{equation*}
\mathbb{E}\left\| U_{0}^{v}\right\| \leq \mathbb{E}\left\| Q_{0}^{v}\right\|
\left\| \widehat{e^{-iv\mathcal{H}}}\right\| \mathbb{E}\left\|
P_{0}^{v}\right\| \leq e^{K_{1}^{2}\mu _{1}v}e^{K_{2}^{2}\mu _{2}v}
\end{equation*}

We shall now prove a similar estimate for the operator $T_{0}^{v}$. Since $%
\left\{ T_{r}^{s}\right\} _{0\leq r\leq s}$ is the resolvent family of
random operators that corresponds to equation (2), the relation $%
T_{0}^{v}=T_{r}^{v}T_{0}^{r}$ holds for all $r\in \left( 0,v\right) $. The
random variables $T_{r}^{v}$ and $T_{0}^{r}$ are independent, and therefore
the inequality

\begin{equation*}
\mathbb{E}\left\| T_{0}^{v}\right\| \leq \mathbb{E}\left\| T_{r}^{v}\right\| 
\mathbb{E}\left\| T_{0}^{r}\right\| 
\end{equation*}
holds. Since the distribution of $T_{r}^{v}$ depends only on $v-r$, the
function $f$ determined by the formula $f(v-r)=\mathbb{E}\left\|
T_{0}^{v}\right\| $ for all $r$ and $v$, 0$\leq r\leq v$, is well defined.

This inequality can be rewritten in terms of $f$:

\begin{equation*}
f(v)\leq f(r)f(v-r).
\end{equation*}

In a similar way, the inequality $f(v)\leq (f(v/n))^{n}$ can be established
for all $v>0$ and $n\in \mathbb{N}$. It follows from the hypotheses of the
lemma that $f^{\prime }(0)$ exists and $f\prime (0)<$1. Consequently, the
relation 
\begin{equation*}
f\left( v\right) \leq \lim_{n\rightarrow \infty }(f(v/n))^{n}=e^{f^{\prime
}\left( 0\right) v}
\end{equation*}
holds. The lemma is proved.

\bigskip

\textbf{Theorem 1} (the randomized Feynman formula). Let $h$ and $k$ be
bounded functions belonging to $L^{2}\left( \mathbb{R}\right) $. Let the map 
$t\mapsto =\mathbb{E}\left\| T_{0}^{v}\right\| $ be differentiable at zero.
Then, for an arbitrary function $\varphi _{0}\in L^{2}\left( \mathbb{R}%
\right) $ and $t>0$, the relation

\begin{equation*}
T_{0}^{t}\varphi _{0}=w-\lim_{n\rightarrow \infty }U_{\left( n-1\right)
t/n}^{t}U_{\left( n-2\right) t/n}^{\left( n-1\right) t/n}\cdots
U_{0}^{t/n}\varphi _{0}
\end{equation*}
holds.

\textit{Proof}. Everywhere below, $t>0$ is arbitrary and fixed and $v=t/n$.
We write the difference between the operators on the left- and right-hand
sides of (14) in the form 
\begin{gather*}
T_{0}^{nv}-U_{\left( n-1\right) v}^{nv}U_{\left( n-2\right) v}^{\left(
n-1\right) v}\cdots U_{0}^{v}=\Pi _{j=n}^{1}T_{\left( j-1\right) v}^{jv}-\Pi
_{j=n}^{1}U_{\left( j-1\right) v}^{jv} \\
=\frac{t}{n}\sum_{j=n}^{1}\Pi _{k=n}^{j+1}T_{\left( k-1\right) v}^{kv}\frac{%
T_{\left( j-1\right) v}^{jv}-U_{\left( j-1\right) v}^{jv}}{v}\Pi
_{k=j-1}^{1}U_{\left( k-1\right) v}^{kv}.
\end{gather*}

For all $v>0$ and $j=1,...,n$ (of course, $v$ and $n$ are related by the
formula $nv=t$), we define $L^{2}\left( \mathbb{R}\right) $-valued random
variables $\xi _{j}^{v}$ by putting 
\begin{equation*}
\xi _{j}^{v}=\Pi _{k=n}^{j+1}T_{\left( k-1\right) v}^{kv}\frac{T_{\left(
j-1\right) v}^{jv}-U_{\left( j-1\right) v}^{jv}}{v}\Pi _{k=j-1}^{1}U_{\left(
k-1\right) v}^{kv}\varphi _{0}.
\end{equation*}
Then the proof of the randomized Feynman formula (14) becomes equivalent to
the verification of the fact that

\begin{equation}
\mathbb{E}\left\| \frac{1}{n}\sum_{j=1}^{n}\xi _{j}^{r}\right\|
_{L^{2}\left( \mathbb{R}\right) }^{2}\rightarrow 0\text{ as }n\rightarrow
\infty .
\end{equation}

To prove (20), it suffices to show that the expressions $\mathbb{E}\left\|
\xi _{j}^{r}\right\| _{L^{2}\left( \mathbb{R}\right) }^{2}$ are uniformly
bounded with respect to $j$ in some neighbourhood of the point $v=0$ and
that the limit relation $\mathbb{E}\left( \xi _{j}^{v},\xi _{k}^{v}\right)
_{L^{2}\left( \mathbb{R}\right) }\rightarrow 0$ as $v\rightarrow 0$ holds
uniformly with respect to all $j\neq k$. For $r,v>0$, let $\mathcal{Y}%
_{r}^{r+v}$ be the minimal 
$\frac34$%
-algebra containing $\mathcal{Q}_{r}^{r+v}$ and $\mathcal{P}_{r}^{r+v}$.

Furthermore, for an arbitrary random variable $\xi $, we denote by $\mathbb{E%
}_{r}^{r+v}\xi $ the conditional mathematical expectation $\mathbb{E}\left[
\xi |\mathcal{Y}_{0}^{r}\cup \mathcal{Y}_{r+v}^{t}\right] $. Thus, $\mathbb{E%
}_{r}^{r+v}$ is the averaging operator over the $\sigma $-algebra $\mathcal{Y%
}_{r}^{r+v}$.

Consider the expression 
\begin{eqnarray*}
\left\| \xi _{j}^{v}\right\| _{L^{2}\left( \mathbb{R}\right) }^{2} &=&\int_{%
\mathbb{R}}\left( \Pi _{k=n}^{j+1}T_{\left( k-1\right) v}^{kv}\frac{%
T_{\left( j-1\right) v}^{jv}-U_{\left( j-1\right) v}^{jv}}{v}\Pi
_{k=j-1}^{1}U_{\left( k-1\right) v}^{kv}\varphi _{0}\right) \left( q\right)
\\
&&\times \overline{\left( \Pi _{k=n}^{j+1}T_{\left( k-1\right) v}^{kv}\frac{%
T_{\left( j-1\right) v}^{jv}-U_{\left( j-1\right) v}^{jv}}{v}\Pi
_{k=j-1}^{1}U_{\left( k-1\right) v}^{kv}\varphi _{0}\right) }\left( q\right)
dq.
\end{eqnarray*}
It follows from the asymptotic expansions (19) and (17) that the relation

\begin{equation*}
\frac{T_{\left( j-1\right) v}^{jv}-U_{\left( j-1\right) v}^{jv}}{v}=\frac{1}{%
v}\sqrt{\mu _{1}\mu _{2}}k\left( \hat{q}\right) h\left( \hat{p}\right)
\Delta W_{1}^{\left( j-1\right) v,v}\Delta W_{2}^{\left( j-1\right)
v,v}+o\left( v^{\frac{1}{2}-\varepsilon }\right)
\end{equation*}
holds. This relation and the estimates for the mathematical expectations of
the operator norms in Lemma 4 imply the inequalities

\begin{eqnarray*}
\left\| \xi _{j}^{v}\right\| _{L^{2}\left( \mathbb{R}\right) }^{2} &=&%
\mathbb{E}\left( \mathbb{E}_{\left( j-1\right) v}^{jv}\left\| \xi
_{j}^{v}\right\| _{L^{2}\left( \mathbb{R}\right) }^{2}\right) \\
&\leq &e^{Cv\left( n-j\right) }[\mu _{1}\mu _{2}\left\| k\right\|
^{2}\left\| h\right\| ^{2}\mathbb{E}\left( \Delta W_{1}^{\left( j-1\right)
v,v}/\sqrt{v}\right) ^{2} \\
&&\times \mathbb{E}\left( \Delta W_{2}^{\left( j-1\right) v,v}/\sqrt{v}%
\right) ^{2}+o\left( v^{1-2\varepsilon }\right) ]e^{2Cv\left( j-1\right)
}\left\| \varphi _{0}\right\| ^{2} \\
&\leq &e^{Ct}[\mu _{1}\mu _{2}\left\| k\right\| ^{2}\left\| h\right\|
^{2}+o\left( v^{1-2\varepsilon }\right) ]\left\| \varphi _{0}\right\| ^{2}.
\end{eqnarray*}
Since $\varepsilon $\TEXTsymbol{>} 0 is arbitrary, the second of these
inequalities provides the desired estimate.

It remains to show that the covariances $\mathbb{E}\left( \xi _{j}^{v},\xi
_{k}^{v}\right) _{L^{2}\left( \mathbb{R}\right) }$ tend to zero as $%
v\rightarrow 0$ uniformly with respect to $j\neq l$. Consider the case $%
j=n,l=n-1$, for which the relation 
\begin{eqnarray*}
\left( \xi _{j}^{v},\xi _{k}^{v}\right) _{L^{2}\left( \mathbb{R}\right) }
&=&\int_{\mathbb{R}}\left( \frac{T_{\left( j-1\right) v}^{jv}-U_{\left(
j-1\right) v}^{jv}}{v}\Pi _{k=j-1}^{1}U_{\left( k-1\right) v}^{kv}\right)
\varphi _{0}\left( q\right) \\
&&\times \left( \overline{T_{\left( n-1\right) v}^{nv}\frac{T_{\left(
n-2\right) v}^{\left( n-1\right) v}-U_{\left( n-2\right) v}^{\left(
n-1\right) v}}{v}\Pi _{k=n-2}^{1}U_{\left( k-1\right) v}^{kv}\varphi _{0}}%
\right) \left( q\right) dq
\end{eqnarray*}
holds.

We first find $\mathbb{E}_{\left( n-1\right) v}^{nv}\left( \xi _{n}^{v},\xi
_{n-1}^{v}\right) _{L^{2}\left( \mathbb{R}\right) }$, that is, we carry out
averaging over the $\sigma $-algebra $\mathcal{Y}_{(n-1)v}^{nv}$. The scalar
product $\left( \xi _{n}^{v},\xi _{n-1}^{v}\right) $ contains only two
random variables depending on $\mathcal{Y}_{(n-1)v}^{nv}$, namely, $\frac{%
T_{\left( n-1\right) v}^{nv}-U_{\left( n-1\right) v}^{nv}}{v}$ and $%
T_{\left( n-1\right) v}^{nv}$.

If $\psi =\Pi _{k=j-1}^{1}U_{\left( k-1\right) v}^{kv}$, then the asymptotic
expansions for these two random variables imply the relation 
\begin{multline*}
\left( \xi _{n}^{v},\xi _{n-1}^{v}\right) _{L^{2}\left( \mathbb{R}\right)
}=([1-\sqrt{\mu _{1}}k\left( \hat{q}\right) \Delta W_{1}^{r,v}-\frac{\mu _{1}%
}{2}vk^{2}\left( \hat{q}\right) +\frac{\mu _{1}}{2}vk^{2}\left( \hat{q}%
\right) \zeta ^{r,v} \\
-\sqrt{\mu _{2}}h\left( \hat{p}\right) \Delta W_{2}^{r,v}-\frac{\mu _{2}}{2}%
vh^{2}\left( \hat{p}\right) +\frac{\mu _{2}}{2}vh^{2}\left( \hat{p}\right)
\eta ^{r,v} \\
-iv\left( k_{0}\left( \hat{q}\right) +h_{0}\left( \hat{p}\right) +\hat{l}%
\right) +o\left( v^{\frac{3}{2}-\varepsilon }\right) ]\frac{T_{\left(
n-2\right) v}^{\left( n-1\right) v}-U_{\left( n-2\right) v}^{\left(
n-1\right) v}}{v}\psi , \\
\left[ \frac{1}{v}\sqrt{\mu _{1}\mu _{2}}k\left( \hat{q}\right) h\left( \hat{%
p}\right) \Delta W_{1}^{\left( j-1\right) v,v}\Delta W_{2}^{\left(
j-1\right) v,v}+o\left( v^{\frac{1}{2}-\varepsilon }\right) \right]
U_{\left( n-2\right) v}^{\left( n-1\right) v}\psi )_{L_{2}\left( \mathbb{R}%
\right) }.
\end{multline*}
To calculate $\mathbb{E}_{\left( n-1\right) v}^{nv}\left( \xi _{n}^{v},\xi
_{n-1}^{v}\right) _{L^{2}\left( \mathbb{R}\right) }$, it is necessary to
remove the two pairs of square brackets in the above relation. It should be
noted that the mathematical expectations of some of the terms resulting from
the removal of the square brackets are zero. Therefore 
\begin{multline*}
\mathbb{E}_{\left( n-1\right) v}^{nv}\left( \xi _{n}^{v},\xi
_{n-1}^{v}\right) _{L^{2}\left( \mathbb{R}\right) }=\mathbb{E}_{\left(
n-1\right) v}^{nv}([1-\sqrt{\mu _{1}}k\left( \hat{q}\right) \Delta
W_{1}^{r,v}-\frac{\mu _{1}}{2}vk^{2}\left( \hat{q}\right) \\
+\frac{\mu _{1}}{2}vk^{2}\left( \hat{q}\right) \zeta ^{r,v}-\sqrt{\mu _{2}}%
h\left( \hat{p}\right) \Delta W_{2}^{r,v}-\frac{\mu _{2}}{2}vh^{2}\left( 
\hat{p}\right) +\frac{\mu _{2}}{2}vh^{2}\left( \hat{p}\right) \eta ^{r,v} \\
-iv\left( k_{0}\left( \hat{q}\right) +h_{0}\left( \hat{p}\right) +\hat{l}%
\right) +o\left( v^{\frac{3}{2}-\varepsilon }\right) ]\frac{T_{\left(
n-2\right) v}^{\left( n-1\right) v}-U_{\left( n-2\right) v}^{\left(
n-1\right) v}}{v}\psi ,o\left( v^{\frac{1}{2}-\varepsilon }\right) U_{\left(
n-2\right) v}^{\left( n-1\right) v}\psi )_{L_{2}\left( \mathbb{R}\right) } \\
+(\left( 1+o\left( 1\right) \right) \frac{T_{\left( n-2\right) v}^{\left(
n-1\right) v}-U_{\left( n-2\right) v}^{\left( n-1\right) v}}{v}\psi ,o\left(
v^{\frac{1}{2}-\varepsilon }\right) U_{\left( n-2\right) v}^{\left(
n-1\right) v}\psi )_{L_{2}\left( \mathbb{R}\right) } \\
+\left( o\left( v^{\frac{3}{2}-\varepsilon }\right) \frac{T_{\left(
n-2\right) v}^{\left( n-1\right) v}-U_{\left( n-2\right) v}^{\left(
n-1\right) v}}{v}\psi ,\sqrt{\mu _{1}\mu _{2}}k\left( \hat{q}\right) h\left( 
\hat{p}\right) U_{\left( n-2\right) v}^{\left( n-1\right) v}\psi \right)
_{L_{2}\left( \mathbb{R}\right) }.
\end{multline*}
It remains to note that $\left\| \psi \right\| \leq e^{C\left( n-2\right)
v}\left\| \varphi _{0}\right\| \leq e^{Ct}\left\| \varphi _{0}\right\| $,
whence it follows that the relation

\begin{gather*}
|\mathbb{E}\left( \xi _{n}^{v},\xi _{n-1}^{v}\right) |=|\mathbb{E}\left\{ 
\mathbb{E}_{\left( n-2\right) v}^{(n-1)v}\mathbb{E}_{\left( n-1\right)
v}^{nv}\left( \xi _{n}^{v},\xi _{n-1}^{v}\right) \right\} | \\
\leq \left( 1+o\left( 1\right) \right) o\left( v^{\frac{1}{2}-\varepsilon
}\right) e^{2Ct}\left\| \varphi _{0}\right\| ^{2} \\
+\left( 1+o\left( 1\right) \right) o\left( v^{\frac{3}{2}-\varepsilon
}\right) \left\| k\right\| \left\| h\right\| \sqrt{\mu _{1}\mu _{2}}%
e^{2Ct}\left\| \varphi _{0}\right\| ^{2}\rightarrow 0
\end{gather*}
holds as $v\rightarrow 0$.

The situation with arbitrary $j$ and $l$ is treated in a similar way. In
this case, the convergence of $\mathbb{E}\left( \xi _{j}^{v},\xi
_{l}^{v}\right) $ to zero is uniform with respect to $j\neq l$ since the
asymptotic expansions for the operators $T_{\left( j-1\right) v}^{jv}$ and $%
U_{\left( j-1\right) v}^{jv}$ as $v\rightarrow 0$ do not depend on $j$.

\section{Stochastic Feynman path integrals over trajectories in the phase
space}

If a function on the phase space depends on a random parameter, then the
Hamiltonian Feynman integral of that function can be defined in a natural way

by extending the classical definition in [24]. Let $(\Omega ,\mathcal{G},%
\mathbb{P})$ be a probability space. Everywhere below, the random Feynman
path integral of a function $F:C([0,t],\mathbb{R})\times \ C([0,t],\mathbb{R}%
)\times \Omega \mapsto \mathbb{C}$ such that $F\left( \xi _{p},xi_{p},\cdot
\right) $ is measurable relative to $(\Omega ,\mathcal{G})$ for all $\xi
_{q} $,$\xi _{p}\in C([0,t],\mathbb{R})$ over trajectories in the phase
space is a random function in $L^{2}\left( \mathbb{R}\right) $ equal to the $%
w$-limit as $n\rightarrow \infty $ of the random functions $z\mapsto
I_{n}\left( F,z\right) $ defined by equation (3).

\bigskip

\textbf{Theorem 2.} Let $h$ and $k$ be bounded real-valued functions
belonging to $L^{2}\left( \mathbb{R}\right) $. Let $h_{0}$ and $k_{0}$ be
functions from $\mathbb{R}$ to $\mathbb{R}$, let $l\in L^{2}\left( \mathbb{R}%
\right) $ be a real-valued function and let $\mathcal{H}%
(q,p)=k0(q)+h0(p)+l(q,p)$ for all $q,p\in \mathbb{R}$. Furthermore, it is
assumed that if $\left\{ T_{r}^{t}\right\} _{0\leq r\leq t}$ is the
resolvent operator family corresponding to equation (2) with Hamiltonian $%
\mathcal{\hat{H}}$ obtained by the qp-quantization of $\mathcal{H}$, then
the map $t\mapsto \mathbb{E}\left\| T_{0}^{t}\right\| $ is differentiable at
zero. Under these conditions, the relation

\begin{eqnarray}
T_{0}^{t}\varphi _{0} &=&\int \exp \left\{ -\int_{0}^{t}\left( i\mathcal{H}%
\left( \xi _{q}\left( s\right) ,\xi _{p}\left( s\right) \right) +\mu
_{1}k^{2}\left( \xi _{q}\left( s\right) \right) +\mu _{2}h^{2}\left( \xi
_{p}\left( s\right) \right) \right) ds\right\}  \notag \\
&&\times \exp \left\{ -\sqrt{\mu _{1}}\int_{0}^{t}k\left( \xi _{q}\left(
s\right) \right) dW_{1}\left( s\right) -\sqrt{\mu _{2}}\int_{0}^{t}h\left(
\xi _{p}\left( s\right) \right) dW_{2}\left( s\right) \right\}  \notag \\
&&\times \varphi _{0}\left( \xi _{q}\left( 0\right) \right) \Phi ^{0,t,\cdot
}\left( d\xi _{q},d\xi _{p}\right)
\end{eqnarray}
holds for an arbitrary function $\varphi _{0}\in L^{2}\left( \mathbb{R}%
\right) $ and all values of $t>0$.

\textit{Proof.} By the definition of the stochastic Hamiltonian Feynman
integral, the right-hand side of (21) is equal to $w-\lim_{n\rightarrow
\infty }B_{t\left( n-1\right) /n}^{t}B_{t\left( n-2\right) /n}^{t\left(
n-1\right) /n}\cdots B_{0}^{t/n}\varphi _{0}$, where for all $r,v>0$ the
action of the operator $B_{r}^{r+v}$ on a function $\varphi \in L^{2}\left( 
\mathbb{R}\right) $ is defined as follows. Let 
\begin{eqnarray*}
f\left( q,p\right)  &=&\exp \{-\left( i\mathcal{H}\left( q,p\right) +\mu
_{1}k^{2}\left( q\right) +\mu _{2}h^{2}\left( p\right) \right) v \\
&&-\sqrt{\mu _{1}}k\left( q\right) \Delta W_{1}^{r,v}-\sqrt{\mu _{2}}%
\int_{0}^{t}h\left( p\right) \Delta W_{2}^{r,v}\}
\end{eqnarray*}
for all $q,p\in \mathbb{R}$. Then $B_{r}^{r+v}\varphi =\hat{f}\varphi $,
where $\wedge $ is the operation of qp-quantization.

We claim that $B_{r}^{r+v}$ coincides with $U_{r}^{r+v}$. Let $f_{1}\left(
q,p\right) =\exp \{\mu _{1}k^{2}\left( q\right) v-\sqrt{\mu _{1}}k\left(
q\right) \Delta W_{1}^{r,v}\}$, $f_{2}\left( q,p\right) =\exp \left\{ -iv%
\mathcal{H}\left( q,p\right) \right\} $ and $f_{3}\left( q,p\right) =\exp
\{\mu _{2}h^{2}\left( p\right) v-\sqrt{\mu _{2}}\int_{0}^{t}h\left( p\right)
\Delta W_{2}^{r,v}\}$ for all $q,p\in \mathbb{R}$. Then, by the definitions
of the corresponding operator families, we have the relations $U_{r}^{r+v}=%
\hat{f}_{1}\hat{f}_{2}\hat{f}_{3}$ and $B_{r}^{r+v}=\widehat{f_{1}f_{2}f_{3}}
$. The second of these was proved in [34] under the condition that $f_{1}$, $%
f_{2}$ and $f_{3}$ are non-random functions. The proof in the stochastic
case is similar.

For every function$g\in L^{2}(\mathbb{R}\times \ L^{2}\left( \mathbb{R}%
\right) )$, we define an operator $\mathcal{J}(g):L^{2}\left( \mathbb{R}%
\right) \mapsto L^{2}\left( \mathbb{R}\right) $ by putting

\begin{equation*}
\lbrack \mathcal{J}(g)\varphi ](q)=\frac{1}{\sqrt{2\pi }}\int_{\mathbb{R}%
}g\left( q,p\right) e^{iqp}\varphi \left( p\right) dp.
\end{equation*}
Then the qp-quantization of $g$ can be written as 
\begin{equation*}
\hat{g}=\mathcal{J}\left( g\right) F^{-1}.
\end{equation*}
The relations 
\begin{equation*}
\widehat{f_{1}f_{2}f_{3}}=\mathcal{J}\left( f_{1}f_{2}f_{3}\right) F^{-1}=%
\hat{f}_{1}\mathcal{J}\left( f_{2}\right) F^{-1}\hat{f}_{3}=\hat{f}_{1}\hat{f%
}_{2}\hat{f}_{3}
\end{equation*}
follow from the fact that $f_{1}$ ($f_{3}$) does not depend on $p$ ($q$),
respectively.

Thus, the relation $B_{v}^{r+v}=U_{v}^{r+v}$ holds for all $r,v>0$, and (21)
follows from Theorem 1.

\bigskip

\textbf{Remark 2.} The result in Theorem 2 can be extended to the case of $%
\tau $-quantization for an arbitrary $\tau \in $[0, 1]. For this, it
suffices to find asymptotic expansions as $v\rightarrow 0$ for the operator $%
B_{v}^{r+v}=\hat{f}$ in the case when $\wedge $ is equal to the operator of $%
\tau $-quantization. It can be shown that these expansions will have the
same form as (17). (Of course, in this case, the operators $\wedge $ on the
right-hand side of this equation will correspond to $\tau $-quantization.)
Then an analogue of Theorem 1 can be proved which will imply the
representation of the solution using the Hamiltonian Feynman integral. The
resulting formula for this representation will coincide with (21) except
that the symbol $\Phi ^{0,t,\cdot }$ is replaced by the pseudo-measure $\Phi
^{t\tau ,t,\cdot }$ corresponding to the case of $\tau $-quantization.

We also note that the use of methods based on Chernoff's theorem in the
stochastic case also leads to randomized analogues of formulae in [9], [33].
For this, suitable expansions into Dyson series must be considered for
random measures depending on realizations of the Wiener processes $W_{1}$
and $W_{2}$.

\textbf{References}

\bigskip

[1] A. L. Alimov, The connection between functional integrals and
differential equations, Teoret. Mat. Fiz. 11:2 (1972), 182190; English
transl. in Theoret. and Math. Phys. 11 (1972).

[2] S. Albeverio and O. G. Smolyanov, Infinite-dimensional stochastic Schr%
\"{o}dinger Belavkin equations, Uspekhi Mat. Nauk 52:4 (1997), 197198;
English transl., Russian Math. Surveys 52:4 (1997), 822823.

[3] V. P. Belavkin and O. G. Smolyanov, The Feynman path integral
corresponding to the stochastic Schr\"{o}dinger equation, Dokl. Akad. Nauk
360 (1998), 589593; English transl., Dokl. Math. 57 (1998), 430434.

[4] J. Gough, Noncommutative Markov approximations, Dokl. Akad. Nauk 379
(2001), 730734; English transl., Dokl. Math. 64 (2001), 112116.

[5] J. Gough, Spectral decomposition of wide-sense stationary quantum random
processes, Dokl. Akad. Nauk 397 (2004), 602605; English transl., Dokl. Math.
70 (2004), 644647.

[6] M. A. Evgrafov, A certain formula for the representation of the
fundamental solution of a differential equation by a continual integral,
Dokl. Akad. Nauk SSSR 191 (1970), 979982; English transl., Soviet Math.
Dokl. 11 (1970), 474478.

[7] V. P. Maslov, Complex Markov chains and the Feynman path integral for
nonlinear equations, Nauka, Moscow 1976. (Russian)

[8] O. G. Smolyanov and A. Truman, Schr\"{o}dinger Belavkin equations and
associated Kolmogorov and Lindblad equations, Teoret. Mat. Fiz. 120:2
(1999), 193207; English transl. Theoret. and Math. Phys. 120 (1999), 973984.

[9] O. G. Smolyanov and E. T. Shavgulidze, Path integrals, Mosk. Gos. Univ.,
Moscow 1990. (Russian)

[10] A. V. Uglanov, On a construction of the Feynman integral, Dokl. Akad.
Nauk SSSR 243 (1978), 14061409; English transl., Soviet Math. Dokl. 19
(1978), 15031587.

[11] R. P. Feynman and A. P. Hibbs, Quantum mechanics and path integrals,
McGrawHill, Maidenhead 1965; Russian transl., Mir, Moscow 1968.

[12] L. Accardi, Y. G. Lu, and I. Volovich, Quantum theory and its
stochastic limit, Springer-Verlag, Berlin 2002.

[13] S. Albeverio, G. Guatteri, and S. Mazzucchi, A representation of the
Belavkin equation via phase space Feynman path integrals, Infin. Dimens.
Anal. Quantum Probab. Relat. Top. 7 (2004), 507526.

[14] S. Albeverio and R. Hoegh-Krohn, Mathematical theory of Feynman
integrals, Lecture Notes in Math., vol. 523, Springer-Verlag, Berlin 1976.

[15] S. Albeverio, V. N. Kolokoltsov, and O. G. Smolyanov, Repr\'{e}%
sentation des solution de l%
\'{}%
equation de Belavkin pour la mesure quantique par une version rigoureuse de
la formule dint\'{e}gration fonctionelle de Menski, C. R. Acad. Sci. Paris S%
\'{e}r. I Math. 323:6 (1996), 661664.

[16] S. Albeverio, V. N. Kolokoltsov, and O. G. Smolyanov, Quantum
restrictions for continuous observation of an oscillator, Rev. Math. Phys. 9
(1997), 907920.

[17] V. P. Belavkin, Nondemolition measurements, nonlinear filtering and
dynamic programming of quantum stochastic processes, Proc. Bellman
Continuous Workshop, Sophia-Antinopolis, Lect. Notes in Computer and Inform.
Sciences 121 (1988), 245265.

[18] V. P. Belavkin, A new wave equation for a continuous nondemolition
measurement, Phys. Lett. A 140 (1989), 355358.

[19] R. Chernoff, Product formulas, nonlinear semigroups and addition of
unbounded operators, Mem. Amer. Math. Soc. 140 (1974), 1121.

[20] E. B. Davies, Quantum theory of open systems, Academic Press, London
1976.

[21] L. Diosi, Continuous quantum measurement and It\={o} formalism, Phys.
Lett. A 129 (1988), 419423.

[22] P. Exner, Open quantum systems and Feynman integrals, Reidel,
Dordrecht, Boston Lancaster 1985.

[23] R. P. Feynman, Space-time approach to nonrelativistic quantum
mechanics, Rev. Mod. Phys. 20 (1948), 367387.

[24] R. P. Feynman, An operation calculus having application in quantum
electrodynamics, Phys. Rev. 84 (1951), 108128.

[25] G. C. Ghirardi , A. Rimini, and T. Weber, Unified dynamics for
microscopic and macroscopic systems, Phys. Rev. D 34 (1986), 471.

[26] R. L. Hudson and K. R. Parthasarathy, Quantum It\={o}'s formula and
stochastic evolutions, Commun. Math. Phys. 93 (1984), 301323.

[27] N. Ikeda and S. Watanabe, Stochastic differential equations and
diffusion processes, North Holland, Kodansha 1981.

[28] P. L\'{e}vy, Th%
\'{}%
eorie de laddition des variables al\'{e}atoires, Gauthier-Villars, Paris
1937.

[29] H. McKean, Stochastic integrals, Academic Press, New York 1969.

[30] M. B. Mensky, Quantum restrictions for continuous observation of an
oscillator, Phys. Rev. D 20 (1979), 384387.

[31] M. B. Mensky, Continuous quantum measurements and path integrals, IOP
Publishing, Bristol 1993.

[32] O. O. Obrezkov, Stochastic Schr\"{o}dinger-type equation with
two-dimensional white noise, Russ. J. Math. Phys. 9:4 (2002), 446454.

[33] O. G. Smolyanov and E. T. Shavgulidze, Some properties and the
applications of Feynman measures in the phase space, Proc. Fourth Vilnius
Conference 2 (1987), 595608.

[34] O. G. Smolyanov, A. G. Tokarev, and A. Truman, Hamiltonian Feynman path
integrals via the Chernoff formula, J. Math. Phys. 43 (2002), 51615171.

[35] O. G. Smolyanov, H. von Weizs\"{a}cker, and O. Wittich, Brownian motion
on a manifold as limit of stepwise conditioned standard Brownian motions,
Canadian Math. Soc. Conf. Proc. 29 (2000), 589602.

[36] O. G. Smolyanov, H. von Weizs\"{a}cker, and O. Wittich, Chernoffs
theorem and the construction of semigroups, Preprint no. 01-11, Institute of
Biomathematics and Biometrics, GSF-National Research Center for Environment
and Health, Neuherberg 2001.

\end{document}